\documentclass[12pt]{iopart}
\thicklines
\usepackage{cite}
\input epsf.tex
\eqnobysec

\begin{document}

{\vbox{
\rightline{SU-ITP-99-48}
\rightline{hep-th/9911002}}}

\title{The Holographic Principle for General Backgrounds}

\author{Raphael Bousso}

\address{Department of Physics, Stanford University,
    Stanford, California 94305-4060 \\
    bousso@stanford.edu}

\begin{abstract}

  We aim to establish the holographic principle as a universal law,
  rather than a property only of static systems and special
  space-times.  Our covariant formalism yields an upper bound on
  entropy which applies to both open and closed surfaces,
  independently of shape or location.  It reduces to the Bekenstein
  bound whenever the latter is expected to hold, but complements it
  with novel bounds when gravity dominates.  In particular, it remains
  valid in closed FRW cosmologies and in the interior of black holes.
  We give an explicit construction for obtaining holographic screens
  in arbitrary space-times (which need not have a boundary).  This may
  aid the search for non-perturbative definitions of quantum gravity
  in space-times other than AdS.

  Based on a talk given at Strings '99 (Postdam, Germany,
  July 1999). More details, references, and examples can be found in
  Refs.~\cite{Bou99b,Bou99c}.

\end{abstract}

\section{Entropy Bounds, Holographic Principle, and Quantum Gravity}
\label{sec-intro}

How many degrees of freedom are there in nature, at the most
fundamental level?  In other words, how much information is required
to specify a physical system completely?  The world is usually
described in terms of quantum fields living on some curved background.
A Planck scale cutoff divides space into a grid of Planck cubes, each
containing a few degrees of freedom.  Applied to a spatially bounded
system of volume $V$, this reasoning would seem to imply that the
number of degrees of freedom, $N_{\rm dof}$,%
\footnote{Strictly, we define $ N_{\rm dof} $ to be the logarithm of
the number of independent basis elements of the quantum Hilbert
space.}
is of the order of the volume, in Planck units:
\begin{equation}
N_{\rm dof}(V) \sim V.
\label{eq-ndof-v}
\end{equation}
Manifestly, the thermodynamic entropy of a system cannot exceed the
number of degrees of freedom: $S \leq N_{\rm dof}$.  One might expect
maximally disordered systems to saturate the inequality, whence
\begin{equation}
S_{\rm max}(V) \sim V.
\label{eq-smax-v}
\end{equation}

This conclusion, however, will have to be rejected when gravity is
taken into account.  The overwhelming majority of states have so much
energy that the system will be inside its own Schwarzschild radius.
If gravitational stability is demanded, these states will be excluded
and the maximal entropy will be much lower than in
Eq.~(\ref{eq-smax-v}).  Guided by the second law of thermodynamics,
Bekenstein~\cite{Bek81} obtained a quantitative result in 1981.  For
spherically symmetric thermodynamic systems it implies that
\begin{equation}
S(V) \leq A/4,
\label{eq-s-a}
\end{equation}
where $A$ is the surface area of the system in Planck units.  This
result does not depend on the detailed properties of the system and
can thus be applied to any spherical volume $V$ of space in which
gravity is not dominant.  The bound is saturated by the
Bekenstein-Hawking entropy associated with a black hole horizon.  In
other words, no stable spherical system can have a higher entropy than
a black hole of equal size.

The Bekenstein bound does not refer to the degrees of freedom underlying
the entropy, so one might think that it leaves the earlier conclusion,
Eq.~(\ref{eq-ndof-v}), intact.  All expected degrees of freedom, a few
per Planck volume, may well be there; the only problem is that if too
many of them are used for generating entropy (or storing information)
the system collapses under its own gravity.

However, a stronger interpretation was proposed by
't~Hooft~\cite{Tho93} and Susskind~\cite{Sus95}: Degrees of freedom
that cannot be utilized should not be considered to exist.  Thus the
number of independent quantum degrees of freedom contained in a given
spatial volume $V$ is bounded from above by the surface area of the
region:
\begin{equation}
N_{\rm dof}(V) \leq A/4
\label{eq-ndof-a}
\end{equation}
A physical system can be completely specified by data stored on its
boundary without exceeding a density of one bit per Planck area.  In
this sense the world is two-dimensional and not three-dimensional as
in Eq.~(\ref{eq-ndof-v}).  For this reason the conjecture is called
the {\em holographic principle.}

The holographic principle constitutes an enormous reduction in the
complexity of physical systems and has dramatic conceptual
implications.  It is far from obvious in a description of nature in
terms of quantum field theories on curved space.  There ought to be a
formulation of the laws of physics in which Eq.~(\ref{eq-ndof-a}) is
manifest: a {\em holographic theory.}  Indeed, it is widely believed
that quantum gravity {\em must}\/ be formulated as a holographic
theory.  This view has received strong support from the AdS/CFT
duality~\cite{Mal97,GubKle98,Wit98a}, which defines quantum gravity
non-perturbatively in a certain class of space-times and involves only
the physical degrees admitted by
Eq.~(\ref{eq-ndof-a})~\cite{SusWit98}.

\section{Limitations of the original formulation}

The holographic principle, Eq.~(\ref{eq-ndof-a}), was proposed as a
universal property of theories that include gravity.  It is believed
to be necessary for (and perhaps to hold the key to) the formulation
of quantum gravity.  In particular, of course, it should apply to
common classical solutions of Einstein's equation.

The validity of the Bekenstein bound, however, is restricted to
systems of ``limited self-gravity''~\cite{Bek81}, i.e., systems in
which gravity is not the dominant force.  Indeed, it is easy to
violate the bound in regions where gravity is dominant, such as a
collapsing star.  The surface area becomes arbitrarily small, while
the enclosed entropy cannot decrease.  Thus Eq.~(\ref{eq-s-a}) will
not hold.  Another example is a super-horizon size region in a flat
Friedmann-Robertson-Walker (FRW) universe~\cite{FisSus98}.  In this
system gravity is dominant in the sense that the overall dynamics is
dictated by the cosmological expansion.  The entropy density is
constant, and volume grows faster than area.  Therefore, the entropy
contained in a large enough sphere will exceed the bound.

Since $S \leq N_{\rm dof}$, the violation of the entropy bound implies
that the holographic principle does not hold in these examples.  This
means that the principle, as it was stated above, is not a universal
law.

\section{Light-sheets}

The limitations of the original proposals motivate us to seek a new,
more general formulation of the holographic principle and the
associated entropy bound%
\footnote{This differs from a more conservative approach taken in
Refs.~\cite{EasLow99,Ven99,BakRey99,KalLin99,Bru99} which aimed to
establish the range of validity of the Bekenstein bound in certain
cosmological solutions.  Our formalism is able to resolve this
interesting question as a special case (see Sec.~\ref{sec-test}).}%
.  We shall retain the formula $S \leq A/4$ and {\em start by
specifying an arbitrary boundary surface}\/ $A$.  The task is to find
a rule that tells us which entropy we mean by $S$.

\subsection{Follow the light}

\begin{figure}[htb!]
  \hspace{.0\textwidth} \vbox{\epsfxsize=1.0\textwidth
  \epsfbox{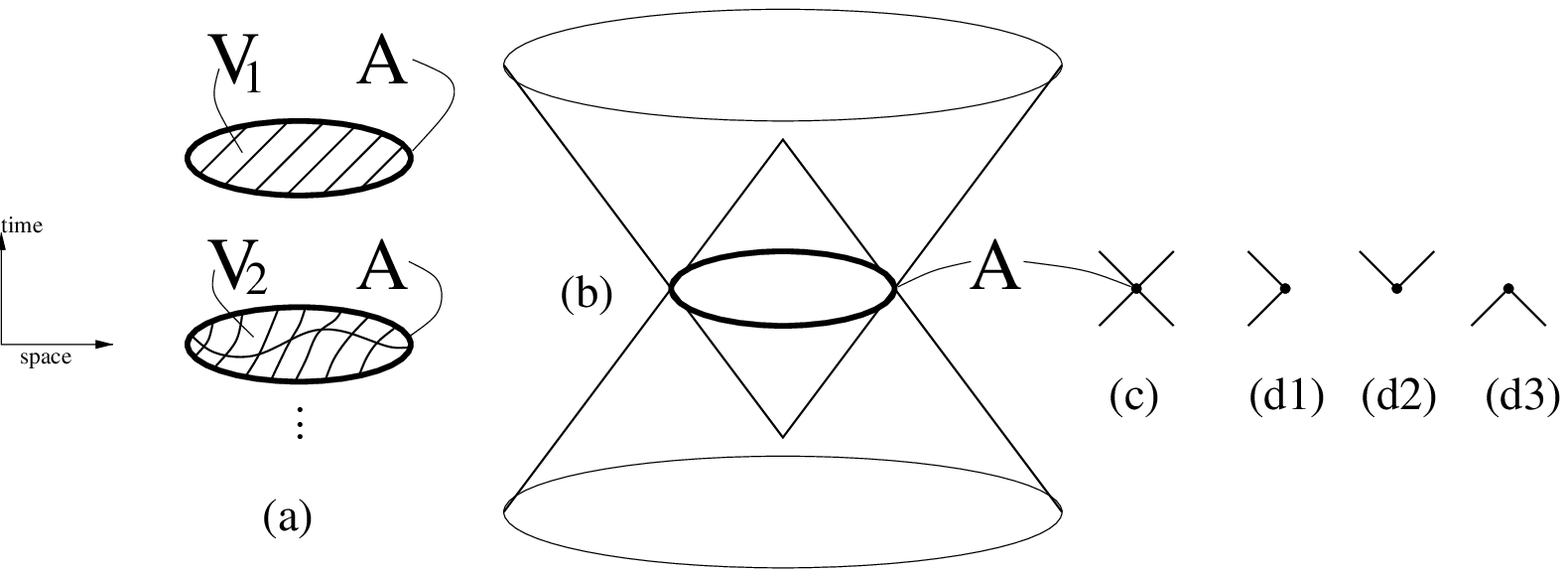}}
\caption%
{\small\sl [In (a,b) we have suppressed one spatial dimension (surface
$\rightarrow$ line).  In (c,d) we have suppressed two (surface
$\rightarrow$ point).  A fixed light-like angle separates spacelike
and timelike directions.]

The spatial volume $V$ enclosed by a surface $A$ depends on
time slicing (a).  Thus the original formulation of the holographic
principle was not covariant.  However, $A$ is the 2D boundary of four
2+1D light-like hypersurfaces (b).  They are covariantly generated by
the past- and future-directed light-rays going to either side of $A$.
E.g., for a normal spherical surface they are given by two cones and
two ``skirts'' (b).

In a Penrose diagram, where spheres are represented by points, the
associated null hypersurfaces show up as the 4 legs of an X (c).  Null
hypersurfaces with decreasing cross-sectional area, such as the two
cones in (b), are called {\em light-sheets}.  The entropy passing
through them cannot exceed $A/4$ (covariant entropy bound).

The light-sheets for normal (d1), trapped (d2), and anti-trapped (d3)
spherical surfaces are shown.  If gravity is weak, as in (b), the
light-sheet directions agree with our intuitive notion of ``inside''
(d1).  For surfaces in a black hole interior, both of the
future-directed hypersurfaces collapse (d2).  Near the big bang, the
cosmological expansion means that the area decreases on both
past-directed hypersurfaces (d3).}
\label{fig-hypersurfaces}
\end{figure}

We shall be guided by a demand for covariance.  Given a closed
surface%
\footnote{By a ``surface'' we mean a surface at some fixed time, i.e.,
a 2-dimensional spatial submanifold.  We do {\em not}\/ mean the
history of a surface (which would be $2+1$-dimensional).} %
of area $A$, the traditional formulation bounds the entropy on the
enclosed volume, or {\em spacelike hypersurface}.  But the
hypersurface we select will depend on the choice of time coordinate
(Fig.~\ref{fig-hypersurfaces}a).  One way to make the statement
covariant would be to demand that the bound hold for {\em all}\/
spacelike hypersurfaces enclosed by $A$.  But this possibility is
already excluded by the counterexamples given above.  Therefore we
must use a {\em null hypersurface}\/ bounded by $A$, following
Fischler and Susskind (FS)~\cite{FisSus98}.  (Several concepts crucial
to a light-like formulation were recognized earlier in
Ref.~\cite{CorJac96}.)

Generalizing from the simple example shown in
Fig.~\ref{fig-hypersurfaces}b, it is easy to see that in fact every
2-dimensional surface, regardless of shape or size, bounds exactly
four null hypersurfaces.  They may be contructed by following the
past- and future-directed orthogonal lightrays leaving the surface on
either side.  But which of the four should be selected for the entropy
bound?  And how far should the light-rays be followed?

\subsection{Go inside}

FS~\cite{FisSus98} considered spherical areas and proposed that the
past-directed {\em ingoing}\/ null hypersurface be used.  The
resulting entropy bound works well for large regions in flat FRW
universes, to which the Bekenstein bound could not be applied.  In
closed or collapsing space-times, however, the FS bound is not
valid~\cite{FisSus98,KalLin99}.  At the root of this difficulty lies
the ambiguity of the concept of ``inside'' in curved, dynamic
space-times.  For a closed surface in asyptotically flat space, the
definition seems obvious: ``Inside'' is the side on which infinity is
not.  But what if space is closed?  For example, which side is inside
the equatorial surface (an $S^2$) of a three-sphere?

\begin{figure}[htb!]
  \hspace{.05\textwidth} \vbox{\epsfxsize=.9\textwidth
  \epsfbox{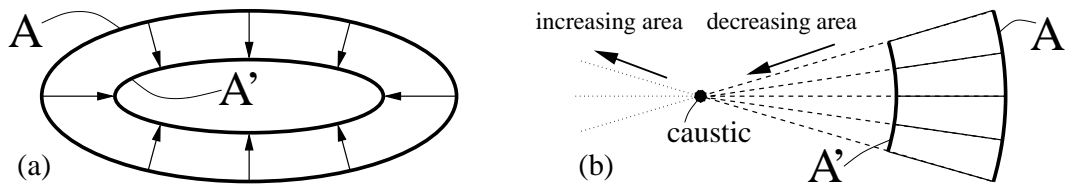}}
\caption%
{\small\sl [Time and one spatial dimension are suppressed.]  We define
the ``inside'' of a 2D surface $A$ to be a light-like direction along
which the cross-sectional area decreases (a): $A' \leq A$, or
equivalently, $\theta \leq 0$.  Such light-rays generate 2+1D {\em
light-sheets}, the entropy on which is bounded by $A/4$.

This definition can be applied to open surfaces as well (b).
Light-sheets end on caustics, as $\theta$ becomes positive there.  If
one stops earlier, the bound can be strengthened to
$(A-A')/4$~\cite{FMW}.}
\label{fig-inside}
\end{figure}

We propose a different definition which is unambiguous, local, and
covariant.  {\em Inside is where the cross-sectional area decreases.}
Consider a two-sphere in flat space (Fig.~\ref{fig-inside}a).  Let us
pretend that we do not know which side is inside.  We can make an
experiment to find out.  We measure the area of the surface.  Now, we
move every point of the surface by some fixed infinitesimal distance
along surface-orthogonal rays (radial rays in this case), to one
particular side.  If this increases the area, it was the outside.  If
the area has decreased, we have gone inside.

It is useful to introduce a slightly more formal language.  The
expansion $\theta$ of a family of light-rays is defined as the
logarithmic derivative of the infinitesimal cross-sectional area
spanned by a bunch of neighbouring light-rays~\cite{HawEll,Wald}.
Thus $\theta$ is positive (negative) if the cross-sectional area is
locally increasing (decreasing).  So the ``inside'' condition is
simply:
\begin{equation}
\theta \leq 0.
\end{equation}
In other words, those light-rays leaving $A$ with negative or zero
expansion generate an ``inside'' null hypersurface bounded by $A$,
which we call a {\em light-sheet}.  The entropy on a light-sheet will
not exceed $A/4$.

If the expansion is positive for the future-directed light-rays to one
side, it will be negative for the past-directed light-rays to the
other side and vice-versa.  Therefore at least two of the four
directions will be allowed.  If the expansion is zero in some
directions, three or even all four null hypersurfaces will be
light-sheets.

It turns out that our covariant definition of ``inside'' has
unexpectedly paved the way for a vast generalization of the formalism
of holography and entropy bounds: In contrast to the naive definition,
there is no need for the surface $A$ to be closed.  It works just as
well for open surfaces, selecting at least two of the four null
hypersurfaces bounded by $A$.  This is illustrated in
Fig.~\ref{fig-inside}b.  In fact the definition is local, so that we
can split a surface into infinitesimal area elements and construct
allowed hypersurfaces piece by piece.  (This permits us to assume,
without loss of generality, that the inside directions are continuous
on $A$; if they flip, we split $A$ into suitable domains.)

\subsection{Know when to stop}

How far do we follow the light-rays?  We need a rule telling us, for
example, to stop at the tip of each of the cones in
Fig.~\ref{fig-hypersurfaces}b.  Otherwise, the light-rays would go on
to generate another cone which could become arbitrarily large.  This
is clearly undesirable.

An elegant and economical feature of the decreasing area rule is that
it has already resolved this question.  We simply insist that $\theta
\leq 0$ {\em everywhere on the null hypersurface}, not only in the
vicinity of $A$.  When neighbouring light-rays intersect, they form a
``caustic'' or ``focal point'' (Fig.~\ref{fig-inside}b).  Before the
caustic, the cross-sectional area is decreasing ($\theta < 0$);
afterwards, it increases and one has $\theta > 0$.  This forces us to
stop whenever a light-ray reaches a caustic, such as the tip of a
light-cone.

\subsection{Summary}

Let $A$ be an arbitrary surface.  We define a {\bf light-sheet} $L(A)$
as {\em a null hypersurface that is bounded by $A$ and constructed by
following a family of light-rays orthogonally away from $A$, such that
the cross-sectional area is everywhere decreasing or constant ($\theta
\leq 0$).}

Since the expansion $\theta$ is a local quantity, the entire
construction is local.  The decreasing area rule enters in two ways:
We use it to determine the ``inside directions;'' as we stressed
above, there will be at least two, and each will yield a distinct
light-sheet.  Having picked one such direction, we then follow each
light-ray no further than to a caustic, where it intersects with
neighbouring light-rays and the area starts to increase.

\section{Covariant Entropy Bound and Holographic Principle}

We are now ready to formulate a {\bf covariant entropy bound}: {\em
Let $A$ be an arbitrary surface area in a physical%
\footnote{This requirement is further discussed in Ref.~\cite{Bou99b},
and in the Appendix.}
space-time, and let $S$ be the entropy contained on any one of its
light-sheets.  Then $S \leq A/4$.}

For spherical surfaces, our conjecture can be viewed as a modification
of the FS bound~\cite{FisSus98}.  It differs in that it considers all
four light-like directions without prejudice and selects some of them
by the criterion of decreasing cross-sectional area.  Unlike previous
proposals, the covariant bound associates entropy-containing regions
with {\em any\/} surface area in {\em any\/} space-time in a precise
way.  Gravity need not be low.  The area can have any shape and need
not be closed.

The covariant bound makes no explicit reference to past and future.
It is manifestly invariant under time-reversal.  In a law about
thermodynamic entropy, this is a mysterious feature.  It cannot be
understood unless we interpret the covariant bound not only as an
entropy bound, but more strongly as a bound on the number of degrees
of freedom that constitute the statistical origin of entropy: $N_{\rm
dof} \leq A/4$, where $N_{\rm dof}$ is the number of degrees of
freedom present on the light-sheet of $A$.  Since no assumptions about
the microscopic properties of matter were made, the limit is
fundamental~\cite{Bou99b}.  There simply cannot be more independent
degrees of freedom on $L$ than $A/4$, in Planck units.  Thus we have
obtained a holographic principle for general space-times%
\footnote{Logically, of course, the entropy bound follows from the
holographic principle, since $S \leq N_{\rm dof}$.  We have found it
interesting to consider the entropy bound first, because the mystery
of its T-invariance leads naturally to the holographic conjecture.
The analogous step had to be considered bold when it was taken by
't~Hooft~\cite{Tho93} and Susskind~\cite{Sus95}, since more
conservative interpretations of the Bekenstein bound were available
(see Sec.~\ref{sec-intro}).  Given the covariant bound, on the other
hand, the necessity for a holographic interpretation is much more
obvious.}%
.

We summarize our conjectures:
\begin{eqnarray}
\mbox{Holographic principle:} & &
 N_{\rm dof}(\mbox{light-sheet}) \leq A/4 \\
\mbox{Covariant entropy bound:~~~~} & &
 S(\mbox{light-sheet}) \leq A/4
\end{eqnarray}
Their non-trivial content lies in the construction of light-sheets
given above.

The bound takes its strongest form when the light-sheet is made as
large as possible, i.e., if we stop only at caustics.  It remains
correct, but becomes less powerful if we choose to stop earlier, as
this will decrease the entropy on the light-sheet, but not the
boundary area.  Flanagan, Marolf and Wald (FMW)~\cite{FMW} have
pointed out, however, that the bound can be strengthened to $(A-A')/4$
in this case.  Here $A'$ is the surface area spanned by the endpoints
of the light-rays (Fig.~\ref{fig-inside}b), which goes to 0 as a
caustic is approached.  This expression is particularly pleasing
because it makes the bound additive over all directions on the
light-sheet, including the transverse (null) direction.  In this form
the covariant entropy bound implies the generalized second law of
thermodynamics~\cite{FMW}.

\section{Evidence for the conjecture}
\label{sec-test}

\subsection{The Bekenstein bound as a special case}

\begin{figure}[htb!]
  \hspace{.2\textwidth} \vbox{\epsfxsize=.6\textwidth
  \epsfbox{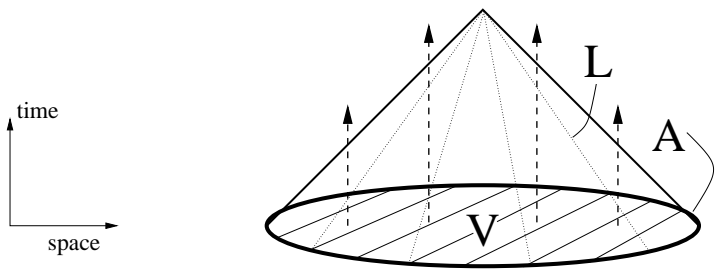}}
\caption%
{The covariant bound implies the Bekenstein bound.  Consider a closed
surface $A$.  We expect the Bekenstein bound to hold only if $A$
encloses a spatial region $V$ of limited self-gravity.  This allows us
to make two assumptions.  1.~$A$ possesses a future-directed
light-sheet $L$ going to the side defined by $V$ (otherwise, $V$ would
contain trapped or anti-trapped surfaces, which implies strong
gravity).  2.~The light-sheet $L$ has no boundary other than $A$,
i.e., it closes off in the center of the enclosed region (otherwise,
the space-time would end within a light-crossing time, which implies
strong gravity).  By causality and the second law, all entropy on the
spatial region $V$ (or more) must pass through the light-sheet; by the
second assumption there are no holes through which it can escape.
Therefore, $S(V) \leq S(L)$.  By the covariant bound, $S(L) \leq A/4$.
Thus we obtain the Bekenstein bound, $S(V) \leq A/4$.}
\label{fig-recover-bek}
\end{figure}

The first important test is whether the covariant bound implies
the Bekenstein bound.  The covariant formulation uses null
hypersurfaces, so how can it bound the entropy on spatial volumes, as
the Bekenstein bound does?  The corresponding argument is presented in
Fig.~\ref{fig-recover-bek}.
It relies on assumptions which can be taken as a definition of
Bekenstein's ``limited self-gravity'' condition.

As an immediate application, we are able to settle a controversial
question which has received much
attention~\cite{EasLow99,Ven99,BakRey99,KalLin99,Bru99}: In
cosmological solutions, what is the largest surface to which
Bekenstein's bound can be reliably applied?  Various types of horizons
were suggested and counter-examples found.  We have established that
the Bekenstein bound holds if the surface permits a complete,
future-directed, ingoing light-sheet.  This singles out the apparent
horizon, which usually separates a normal from an anti-trapped region.
Nevertheless, the claim that the Bekenstein bound holds for the
apparent horizon~\cite{BakRey99} is not always valid~\cite{KalLin99},
since the completeness condition must also be satisfied~\cite{Bou99b}.

While it yields a precise formulation of the Bekenstein bound as a
special case, the covariant entropy bound is more general.  We have
already stressed that it applies also to open surfaces.  Moreover, it
is valid in strongly gravitating regions for which no entropy bounds
were previously available (or even hoped for).  As an illustration of
this claim, let us test the bound for surfaces deep inside a black
hole.

\subsection{Trapped surfaces and a new type of entropy bound}

One can set up a worst-case scenario in which one considers the
horizon surface $A$ of a small black hole at some moment of time.  It
possesses a future-directed light-sheet $L$ which will coincide with
the horizon as long as no additional matter falls in.  Far outside the
black hole, we may set up a highly enthropic shell of arbitrary mass.
In order to avoid angular caustics~\cite{Bou99b}, which would
terminate parts of the light-sheet, we assume exact spherical
symmetry.  Entropy will be carried in radial modes and can be
arbitrarily large.  The shell is allowed to collapse around the small
black hole.  When it reaches the light-sheet, the configuration is
already deep inside a much larger black hole of the shell mass, and no
conventional entropy bound applies.  One might expect a violation of
the bound, since all of the shell will necessarily squeeze through the
radius of the small black hole and reach the singularity at the
center.  It is not difficult to verify, however, that the light-sheet
$L$ ends before any entropy in excess of $A/4$ passes though
it~\cite{Bou99b}.  If the shell entropy exceeds $A/4$, the light-sheet
will reach $r=0$ before all of the shell does.  The bound can be
saturated, but not exceeded.  Thus we see that for trapped surfaces,
the covariant formalism implies genuinely new entropy bounds, which
could not have been anticipated from the original formulation.

\subsection{The closed universe}

\begin{figure}[htb!]
  \hspace{.1\textwidth} \vbox{\epsfxsize=.8\textwidth
  \epsfbox{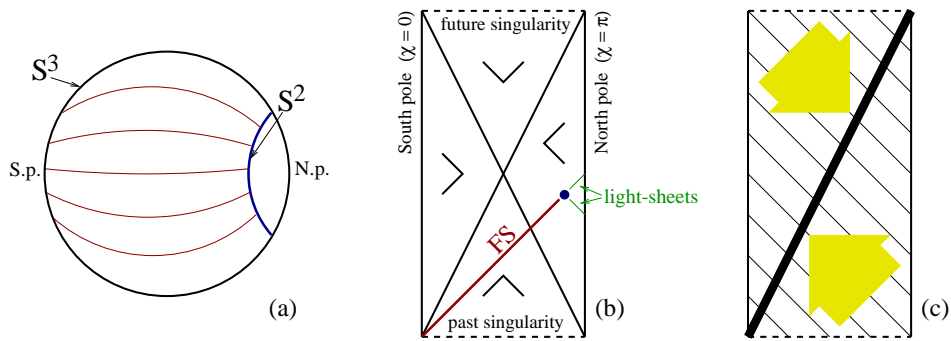}}
\caption%
{\small\sl The closed FRW universe.  A small two-sphere divides the
$S^3$ spacelike sections into two parts (a).  The covariant bound
will select the small part, as indicated by the normal wedges (see
Fig.~\ref{fig-hypersurfaces}d) near the poles in the Penrose diagram
(b).  After slicing the space-time into a stack of light-cones, shown
as thin lines (c), all information can be holographically projected
towards the tips of wedges, onto an embedded screen hypersurface (bold
line).}
\label{fig-closed-all}
\end{figure}

It is particularly instructive to verify the covariant bound for a
closed, matter-dominated FRW universe, to which the FS
bound~\cite{FisSus98} could not be applied.  Consider a small
two-sphere near the turn-around time (Fig.~\ref{fig-closed-all}a).
The FS bound would consider the entropy on a light-cone which
traverses the large part of the $S^3$ space
(Fig.~\ref{fig-closed-all}b).  This would be almost the entire entropy
in the universe and would exceed the arbitrarily small area of the
two-sphere.  The covariant bound, on the other hand, considers the
entropy on light-sheets which are directed towards the smaller part
(the polar cap).  This entropy vanishes as the two-sphere area goes to
zero~\cite{Bou99b}.  This illustrates the power of the decreasing area
rule.

\subsection{Questions of proof}

More details and additional tests are found in Ref.~\cite{Bou99b}.  No
physical counterexample to the covariant entropy bound is known (see
the Appendix).  But can the conjecture be proven?  In contrast with
the Bekenstein bound, the covariant bound remains valid for unstable
systems, for example in the interior of a black hole.  This precludes
any attempt to derive it purely from the second law.  Quite
conversely, the covariant bound can be formulated so as to imply the
generalized second law~\cite{FMW}.

FMW~\cite{FMW} have been able to derive the covariant bound from
either one of two sets of physically reasonable hypotheses about
entropy flux.  In effect, their proof rules out a huge class of
conceivable counterexamples.  Because of the hypothetical nature of
the FMW axioms and their phenomenological description of entropy,
however, the FMW proof does not mean that one can consider the
covariant bound to follow strictly from currently established laws of
physics~\cite{FMW}.  In view of the evidence we suggest that the
covariant holographic principle itself should be regarded as
fundamental.

\section{Where is the boundary?}

Is the world really a hologram~\cite{Sus95}?  The light-sheet
formalism has taught us how to associate entropy with arbitrary 2D
surfaces located anywhere in any spacetime.  But to call a space-time
a hologram, we would like to know whether, and how, {\em all}\/ of its
information (in the entire, global 3+1-dimensional space-time) can be
stored on some surfaces.  For example, an anti-de~Sitter ``world'' is
known to be a hologram~\cite{Mal97,SusWit98}.  By this we mean that
there is a one-parameter family of spatial surfaces (in this case, the
two-sphere at spatial infinity, times time), on which all bulk
information can be stored at a density not exceeding one bit per
Planck area.  Such surfaces will be called {\em screens}\/ of a
spacetime.  Can analogous screens be found in other space-times?

The trick is to slice the 3+1 space-time into a one parameter family
of 2+1 null hypersurfaces $H(\tau)$.  On each null hypersurface
$H(\tau)$, locate the 2D spatial surface $A_{\rm max}(\tau)$ of
maximum area.  Unless it lies on a space-time boundary, this surface
divides $H(\tau)$ into two parts.  But the cross-sectional area
decreases in both directions away from $A_{\rm max}(\tau)$ by
construction.  Thus the entropy on the entire hypersurface $H(\tau)$
cannot exceed $A_{\rm max}(\tau)/2$.  In other words, all the physics
on $H(\tau)$ can be described by data stored on $A_{\rm max}(\tau)$ at
a density not exceeding 1 bit per Planck area.

By repeating this construction for all values of the parameter $\tau$,
one obtains a one parameter family of 2D screens, $A_{\rm max}(\tau)$.
It forms a 2+1 hypersurface $\cal A$, which is embedded in the bulk
space-time and on which the entire space-time information can be
stored holographically.  As an example, Fig.~\ref{fig-closed-all}c
shows the construction of a holographic screen hypersurface in a
closed, matter-dominated FRW universe.

Other examples can be found in Ref.~\cite{Bou99c}.  Note that $\cal A$
will depend on the choice of slicing, and will not always be
connected.  In Minkowski space, one finds that either one of the two
null infinities can play the role of a global screen.  In global
de~Sitter, one must use both past and future infinity; however, the
observable part of de~Sitter can be encoded on the event horizon,
which forms a null hypersurface of finite area.

\section{Dreams of a holographic theory}

The stunning success of the AdS/CFT duality~\cite{Mal97} has led to
speculations that a non-perturbative definition of quantum gravity
would involve theories on holographic screens in other space-times as
well.  Such hypotheses were vague, however, because no general
definition of holographic screens was available.  In particular, it
was completely unclear how holography could be compatible with
space-times that had no boundary, such as a closed FRW universe.

What we have shown is that embedded holographic screens exist in any
space-time, and how to construct them.  This result lends strong
support to the holographic hypothesis.  We hope that it will be of use
in the search for a fully general, manifestly holographic unified
theory.

Indeed, the structure of screens in cosmological solutions provides
interesting constraints on the general formulation of holographic
theories.  Generically, the screen area (and thus the number of
degrees of freedom of a manifestly holographic theory defined on
screens) will be time-dependent.  Moreover, the causal character of
the screen hypersurface $\cal A$ can change repeatedly between
Euclidean and Lorentzian.

These pathologies appear to exclude the possibility of formulating a
well-defined, conventional theory on the screens.  Moreover, in
generalizing a screen theory approach \`a la AdS one would encounter
the basic drawback that the screen itself must be put in by hand, for
example by constraining oneself to asymptotically AdS spaces.
Ideally, all geometric features should come out of the theory.

A background-independent holographic theory is likely to require a
more radical approach.  We suspect that quantum gravity is a
pre-geometric theory containing a sector from which classical general
relativity can be constructed.  The recovery of local physics will
likely be highly non-trivial, as can be gleaned from the example of
AdS/CFT duality.  One would need to build the geometry from a
pre-geometric set of degrees of freedom in such a way that the
covariant holographic principle is manifestly satisfied.  This would
be the fundamental origin of the principle.

\section*{Appendix}

Here we add some comments on the generality of the covariant entropy
bound.  We also discuss energy conditions and respond to recent
criticism.

\mbox{}\\[-2ex] \noindent{\em Classical gravity.~~} The covariant
entropy bound applies to all physical solutions of classical general
relativity.  This constitutes an enormous generalization compared to
previous formulations of entropy bounds~\cite{Bek81,FisSus98}.  We
obviously do not mean to assume that $\hbar$ is exactly zero.  Rather,
the term ``classical'' refers to space-times in which quantum
gravitational effects are negligible.  This holds for most known
solutions, in particular for cosmology on scales larger than the
Planck scale, for ordinary thermodynamic systems, and for black holes
on time-scales less than the evaporation time.  The covariant entropy
bound (and the Bekenstein bound, where applicable) provides a simple,
yet highly non-trivial and powerful relation between area and entropy
in such space-times.  This renders nugatory the observation by
Lowe~\cite{Low99} that the bound, $A/4G\hbar$, diverges for $\hbar
\rightarrow 0$.

\mbox{}\\[-2ex] \noindent{\em Quantum and semi-classical gravity.~~}
The holographic principle may hold a key to the non-perturbative
formulation of string theory and quantum gravity.  This does not mean
that we should expect it to remain useful in the strong quantum
regime, where an approximation to classical geometry need not exist.
Indeed, the very concept of area might become meaningless.  As we
argued above, the fundamental role of the holographic principle would
then be in the recovery of classical relativity from a suitable sector
of a pre-geometric unified theory.  In semi-classical gravity,
however, there ought to be a place for the holographic principle.
Indeed, we are unaware of any semi-classical counter-examples to the
covariant entropy bound.  However, the situation is more subtle than
in classical gravity.  It is easy to construct apparent problems, but
they turn out to stem from limitations of the semi-classical
description, not of the holographic principle.

Consider, for example, a sphere just outside a black hole horizon.
This surface possesses a future-directed light-sheet which crosses the
horizon and sweeps the entire interior of the black hole.  How much
entropy is on the light-sheet?  Naively, we might count first the
Bekenstein-Hawking entropy of the horizon ($A/4$, which by itself
saturates the bound).  In addition, we might count the ordinary
thermodynamic entropy of the matter which formed the black hole and
passes through the interior part of the light-sheet.  This would lead
to a violation.  The breakdown is not in the holographic principle,
however, but in our naive interpretation of the semi-classical
picture.  We have overcounted.  The horizon entropy represents the
potential information content of the system to an observer outside the
black hole.  The matter entropy represents the actual information to
an observer falling in.  The viewpoints are complementary, and it is
operationally meaningless to take a global stance and count both
entropies.

\mbox{}\\[-2ex] \noindent{\em Resolving Lowe's objection.~~}
Lowe~\cite{Low99} has argued that the covariant entropy
bound~\cite{Bou99b} fails for a semi-classical black hole in thermal
equilibrium with a surrounding heatbath.  He considers a surface $A$
on the black hole horizon, which possesses a future-directed outgoing
light-sheet $L$.  Two assumptions are made about this
set-up~\cite{Low99}.  1) $L$ will coincide with the black hole horizon
forever.  2) The configuration is stable and thus eternal.  By the
second assumption, an infinite amount of entropy crosses the black
hole horizon.  By the first assumption, this entropy crosses the
light-sheet $L$ of a finite area $A$.  Thus, it would seem, the bound
is violated.

Because black holes have negative specific heat, however, the
configuration is in fact unstable~\cite{GroPer82}.  The black hole
decays in a runaway process of accretion or evaporation.  This
invalidates Lowe's second assumption and thus his conclusion.  The
static approximation used in Ref.~\cite{Low99} breaks down completely
within a time of the order of the black hole evaporation time-scale,
$M^3$.  During this time, the bound will be satisfied, since no more
than the black hole entropy, $A/4$, gets exchanged in an evaporation
time.  (We thank L.~Susskind for a discussion.)

\mbox{}\\[-2ex] \noindent{\em Can Lowe's criticism be strengthened?~~}
Stable semi-classical black holes actually do exist in AdS space.
However, they do not provide a counter-example either, because the
global structure of AdS does not permit us to pipe information through
a black hole as would be required in order to violate the bound.
Instead, the Hawking radiation forms a thermal atmosphere around the
black hole.  No new entropy can be introduced without increasing the
area of the black hole.

It is interesting to note that Lowe's first assumption is also
problematic.  It does hold, of course, for a classical black hole in
vacuum.  In a semi-classical equilibrium, however, the ingoing
radiation necessarily possesses microscopic fluctuations.  They will
get imprinted on the expansion $\theta$ of the light-sheet generators.
Thus, $\theta$ will not vanish exactly at all times.  The
$\theta^2$-term in Raychauduri's equation effectively provides a bias
towards negative values of $\theta$:
\begin{equation}
\dot{\theta} = - \frac{1}{2} \theta^2 + \mbox{ terms which
average to zero}
\label{eq-nobalance}
\end{equation}
We see that the effects of in- and out-going radiation on the
expansion will not cancel out.  Eq.~(\ref{eq-nobalance}) leads to a
runaway process in which the light-sheet $L$ departs from the apparent
horizon and collapses.

It is not clear whether, in a baroque twist, one could set up a demon
to keep the black hole at constant size.  The demon would have to
maintain a position close enough to the black hole to react to
temperature changes by increasing or decreasing the matter influx, and
it would have to perform this task for a time longer than $M^3$
without violating the self-consistency of the solution.  But even if
we assume that this was possible, and that the black hole could be
kept at a constant size, the light-sheet would collapse due to small
fluctuations.  The black hole horizon might last forever, but a
light-sheet with $\theta \leq 0$ will not.  Thus one might be able to
transfer an infinite amount of entropy through the horizon, but not
through the light-sheet.

\mbox{}\\[-2ex] \noindent{\em Further objections.~~} Lowe also points
out that angular caustics~\cite{Bou99b} cannot protect the bound if
only radial modes carry entropy.  Because we shared this concern,
however, we addressed the issue explicitly in Sec.~6.2 of the
criticized paper~\cite{Bou99b}, where it was shown that the bound can
be saturated, but not exceeded, in the worst-case scenario of
exactly-spherical collapse of an arbitrarily massive shell.  The
various objections raised by Lowe against the Bekenstein bound have
previously been answered by Bekenstein and others (see, e.g.,
Refs.~\cite{Bek83,Bek84,SchBek89,Bek99}).

\mbox{}\\[-2ex] \noindent{\em Conditions on matter.~~} In particular,
we stress that the covariant bound, like the Bekenstein bound, is
conjectured to hold only for matter that actually exists in nature.
It thus predicts that the fundamental theory will not contain an
exponentially large number of light non-interacting
particles~\cite{Bek99} or permit the kind of negative energy densities
that would be needed to break the bound.  These requirements are met
both empirically and by current theoretical models.  In this sense we
conjecture the bound to be generally valid.  We cannot spell out
precise conditions on matter until the fundamental theory is fully
known.  Its matter content is unlikely to satisfy any of the usual
energy conditions on all scales.  However, at the price of making the
conjecture slightly less general than necessary, we can obtain a
testable prediction for a huge class of space-times by demanding the
dominant energy condition~\cite{Bou99b}.

\mbox{}\\[-2ex] \noindent{\em A proposed modification.~~} Instead of
using the $\theta \leq 0$ rule to determine where to end a
light-sheet, Tavakol and Ellis (TE)~\cite{TavEll99} have suggested an
interesting modification%
\footnote{We thank Ted Jacobson for discussions of related ideas.}%
.  They propose to terminate when generating light-rays depart from
the boundary of the causal past or future of $A$.  Basically, this
amounts to the following difference.  In our
prescription~\cite{Bou99b}, one stops only when {\em neighbouring}\/
light-rays intersect locally, i.e., at caustics.  The TE proposal
would be to stop also when {\em non-neighbouring}\/ light-rays
intersect.  This makes no difference in spherically symmetric
situations, but generically it leads to smaller light-sheets and thus
to a weaker bound.  In the absence of any counter-examples to our
formulation~\cite{Bou99b}, and in view of the FMW results~\cite{FMW},
this relaxation does not appear to be necessary.

Moreover, the TE formulation is not self-consistent unless one gives
up one of the most attractive features of the light-sheet formalism:
its locality in the area.  This is most easily seen by considering (in
a nearly empty 2+1D world) a 1D oval ``surface'' $A$ consisting of two
half-circles, $A_1$ and $A_2$, joined by long, parallel line segments
$A_3$ and $A_4$.  Non-neighbouring light-rays intersect on a line
midway between $A_3$ and $A_4$.  If one stopped there, the 1+1D
light-sheet would take the shape of a roof.  Now consider only one of
the parallel line segments, $A_3$, by itself.  If we construct its
light-sheet individually, the light-rays will continue indefinitely
unless they are bent into caustics by some matter.  The entropy on
this large light-sheet will be bounded by $A_3/4$.  Thus, if we
applied the alternative rule~\cite{TavEll99} to each area element
separately, and added up the resulting bounds, we would recover the
covariant entropy bound in our formulation~\cite{Bou99b}.

Tavakol and Ellis~\cite{TavEll99} correctly point out that
light-sheets can be extremely complicated structures in inhomogeneous
space-times.  But this should not motivate us to change the $\theta
\leq 0$ rule, just as we would not discard the standard model merely
because its application becomes impractically complicated when one is
describing an elephant.  Of course, it may often be practical to
consider the smaller light-sheets suggested by TE.  But we are
interested mostly in the fundamental theoretical role of the
holographic principle.  Therefore we advocate its strongest and most
general formulation~\cite{Bou99b,Bou99c,FMW}.

\section*{Acknowledgments}

Since Refs.~\cite{Bou99b,Bou99c} appeared, I have enjoyed stimulating
discussions and useful correspondence with many friends and
colleagues, including Nima Arkani-Hamed, Vijay Balasubramanian, Jacob
Bekenstein, Ramy Brustein, Sean Carroll, Andrew Chamblin, \'Eanna
Flanagan, Edi Halyo, Stephen Hawking, Gary Horowitz, Ted Jacobson,
Nemanja Kaloper, Andrei Linde, Juan Maldacena, Don Marolf, Amanda
Peet, Simon Ross, Steve Shenker, Lee Smolin, Andy Strominger, Lenny
Susskind, Reza Tavakol, Gabriele Veneziano, Bob Wald, and Ed Witten.

I would like to thank the organizers of Strings '99 for a stimulating
meeting.

\bigskip\bigskip
\bigskip\bigskip
\section*{References}


\end{document}